\documentclass[prl,aps,twocolumn,superscriptaddress]{revtex4-1}

\usepackage{bm}
\usepackage[colorlinks=true,linkcolor=blue,citecolor=blue]{hyperref}
\usepackage{times}
\usepackage{amsmath}
\usepackage{amssymb}
\usepackage{amsthm}
\usepackage{amsfonts}
\usepackage{enumerate}
\usepackage{latexsym}
\usepackage{ifpdf}
\newcommand{\beq}{\begin{equation}}
\newcommand{\eeq}{\end{equation}}
\usepackage{graphicx}
\usepackage{makeidx}
\usepackage{color}

\begin{document}

\title{Disentangled cooperative orderings in artificial rare-earth nickelates}

\author {S. Middey}
\altaffiliation{Contributed equally}
\email{smiddey@iisc.ac.in }
\affiliation{Department of Physics, Indian Institute of Science, Bangalore 560012, India}
\author{D. Meyers }
\altaffiliation{Contributed equally}
\email{dmeyers@bnl.gov}
\affiliation{Department of Condensed Matter Physics and Materials Science, Brookhaven National Laboratory, Upton, New York 11973, USA}
\author{M. Kareev}
\affiliation{Department of Physics and Astronomy, Rutgers University, Piscataway, New Jersey 08854, USA}
\author{Y. Cao}
\affiliation{Department of Physics and Astronomy, Rutgers University, Piscataway, New Jersey 08854, USA}
\author{X. Liu}
\affiliation{Department of Physics and Astronomy, Rutgers University, Piscataway, New Jersey 08854, USA}
\author{P. Shafer}
\affiliation {Advanced Light Source, Lawrence Berkeley National Laboratory, Berkeley, California 94720, USA}
\author{J. W. Freeland}
\affiliation {Advanced Photon Source, Argonne National Laboratory, Argonne, Illinois 60439, USA}
\author{J. W. Kim}
\affiliation {Advanced Photon Source, Argonne National Laboratory, Argonne, Illinois 60439, USA}
\author{P. J. Ryan}
\affiliation {Advanced Photon Source, Argonne National Laboratory, Argonne, Illinois 60439, USA}
\author{ J. Chakhalian}
\affiliation{Department of Physics and Astronomy, Rutgers University, Piscataway, New Jersey 08854, USA}

\begin{abstract}
Coupled transitions between distinct ordered phases are important aspects behind the rich phase complexity of correlated oxides that  hinders our understanding of the underlying phenomena. For this reason, fundamental control over  complex transitions  has  become a leading motivation  of the  designer approach  to materials. We have devised a series of new superlattices by combining a Mott insulator and a correlated metal to form ultra-short period superlattices, which allow one to disentangle the simultaneous orderings in $RE$NiO$_3$.  Tailoring an incommensurate heterostructure period relative to the bulk charge ordering pattern suppresses the charge order transition while preserving metal-insulator and antiferromagnetic transitions.  Such selective decoupling of the entangled phases resolves the long-standing puzzle about the driving force behind the metal-insulator transition  and points to the site selective Mott transition as the operative mechanism. This designer approach  emphasizes the  potential of heterointerfaces for selective control of simultaneous transitions in complex materials with entwined broken symmetries.

 \end{abstract}
 
\maketitle

Materials exhibiting complex phase diagrams are often characterized by multiple entangled order parameters.  Complex oxides in particular show multiple types of   phase transformations controlled by various  external stimuli~\cite{mit_rmp,pressure,jak_rmp,vo2_ediffraction,tokura_manganites,pnictide,kiemer_nature}. However, understanding of the mechanism behind the transitions is quite challenging in compounds which display multiple simultaneous orderings such as cuprates, vanadates, manganites,  pnictides,  etc.~\cite{vo2_ediffraction,tokura_manganites,pnictide,kiemer_nature}.   In recent years substantial efforts have been put forward  to  decouple  structural and electronic transitions by external means. For example,  these  transitions in vanadates  can be separated  by external stimuli~\cite{v2o3_pressure,vo2_ediffraction} leading to the discovery  of a novel monoclinic metallic phase. This  surprising finding has triggered the search for alternative methods of  materials design in hopes of realizing such unusual phases without external perturbations.  In the rapidly progressing field of controllable tunability in complex oxides, a promising emerging method is the heterostructuring of ultra-thin dissimilar layers into a superlattice  without bulk analogues~\cite{jak_rmp}. Such artificial structures allow layer-by-layer design and the exploration of unique stacking sequences which  offer an unprecedented opportunity for manipulation of materials properties.

 \begin{figure}  
\vspace{-0pt}
\includegraphics[width=.45\textwidth] {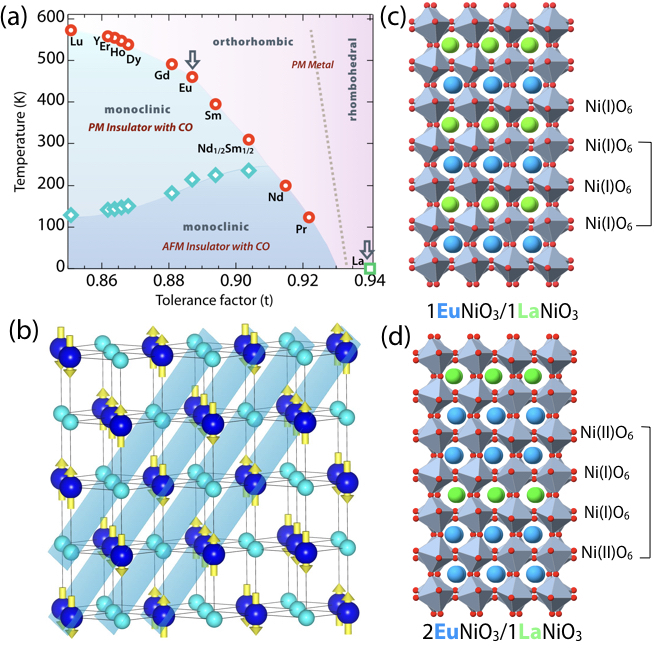}
\caption{\label{}  (a) Phase diagram of bulk $RE$NiO$_3$~\cite{rno_phase1,rno_phase,own_review}. Arrows denote individual members, which are used in this paper to make superlattices.  (b) Rock-salt type charge ordering with exaggerated Ni$^{3\pm \delta}$ radius variations and E$^\prime$-AFM spin ordering ~\cite{haule} for bulk $RE$NiO$_3$ ($RE$ and O atoms omitted for clarity). Pseudocubic (pc) (1 1 1) planes are highlighted. Schematic crystal structure of (c) 1ENO/1LNO and (d)  2ENO/1LNO  superlattices grown along pseudo cubic [0 0 1] direction. All Ni's in 1ENO/1LNO SL have Eu in one side and La on the opposite side along [0 0 1]$_\mathrm{pc}$. However, 2ENO/1LNO has two types of Ni: NiI with Eu and La on opposite sides and NiII with Eu on both sides.}
\end{figure}

Another example of a complex multi-order parameter system,  $RE$NiO$_3$ ($RE$ = La, Pr, Nd, Sm, Eu, ...Lu etc.)~\cite{rno_phase1,rno_phase}  have recently undergone intense scrutiny due to their simultaneous metal-to-insulator (MIT), structural,  charge-ordering  (CO)/bond disproportionation (BD), and unusual E$^{\prime}$-type antiferromagnetic (E$^{\prime}$-AFM) transitions~\cite{own_review}. The combination of multiple transitions occurring at the same temperature (Fig. 1(a)) has fueled an active debate on the origin of the MIT. Among several mechanisms, the CO  on Ni$^{3+}$ with a rock-salt pattern, i.e. $d^7$ + $d^7$ $\leftrightarrow$ $d^{{7}+\delta}$ + $d^{7-\delta}$, (Fig. 1(b)) has been fastidiously scrutinized by various probes and  shown to accompany the MIT both in bulk and thick films~\cite{nno_co,scagnoliprb,lorenzoprb,keimer_scattering} with simultaneous lattice symmetry lowering from orthorhombic to monoclinic. While density functional theory  supports the scenario of a CO-driven MIT~\cite{Mazin},  recent infrared spectroscopy measurements strongly emphasized the importance of Mott physics~\cite{Stewart}. However, charge disproportionation is not required for a  pure Mott-type MIT ~\cite{nno_co,scagnoliprb,lorenzoprb,keimer_scattering,Upton,derek_nno,keimer_raman} and the expected $T_{\textrm{MIT}}$  would have been well above $T_{N}$ (N\'{e}el temperature) irrespective of the choice of a $RE$ ion, which clearly contradicts experiments. Recent theories~\cite{khomskii_holeordering,millis_siteselective,sawatzky_hf,georges_dmft,haule,green} have instead  postulated that this type of MIT is a Ni `site selective Mott' (SSM) transition  and occurs without any explicit charge ordering on the Ni sites, i.e. $d^8\underline L +d^8\underline L \leftrightarrow d^8$ ($S$=1) + $d^8\underline L^2$ ($S$=0), where $\underline L$ indicates a hole on  oxygen $p$ orbitals~\cite{fujimori}. The insulating phase obtained by the SSM transition does not have conventional charge disproportionation Ni$^{+3 \pm \delta}$ and  instead the theory  implies a  `bond disproportionation' (BD) phase as different Ni sites with  $d^8$  and $d^8\underline L^2$ configuration will have longer and shorter Ni-O bond-length respectively. In sharp  contrast to  SSM, the appearance of the insulating phase with simultaneous E$^{\prime}$-type antiferromagnetic (E$^\prime$-AFM) ordering $T_\mathrm{MIT}$ = $T_N$   for  NdNiO$_3$ and PrNiO$_3$ has been explained by Fermi surface nesting~\cite{balents_landau,balents_hf,Frano_prl}. Further, E$^\prime$-AFM transition has been also reported in the weakly metallic state of PrNiO$_3$/PrLaO$_3$ superlattices~\cite{keimer_raman}  without or with very weak charge ordering and also in some metallic $RE$NiO$_3$~\cite{enopressure,lnoafm}.  All these conflicting results highlight the need for  an alternative approach to materials design  to selectively decouple the entangled orderings and elucidate the individual role of each  for the MIT.

In this letter, we utilize a new  route to control  the transitions by mapping the number of dissimilar atomic planes of the heterostructure to  the anticipated real-space  pattern for a specific ordering. Two different members of the series,  insulating EuNiO$_3$ (ENO)~\cite{derek_prb,derek_prb2}  and correlated metal LaNiO$_3$ (LNO)~\cite{lno_optics} have been heterostructured  to  stabilize  a new quantum material: [\textit m u.c. EuNiO$_3$ / \textit n u.c. LaNiO$_3$]  (Fig. 1(c), (d))   where \textit m=1, 2 and \textit n=1  refer to the individual layer thickness (u.c. = unit cell in pseudo-cubic notation).  While $RE$NiO$_3$ SLs encompassing band-insulating spacing layers are ubiquitous~\cite{own_review}, our approach of combining two distinct $RE$NiO$_3$ layers in the form of short-period superlattices  has not been explored so far to the best of our knowledge.
The design idea is illustrated in Fig. 1(c), (d). All Ni sites  are structurally equivalent in 1ENO/1LNO superlattice (SL)  and the checker board type CO  can be naturally accommodated within this structure, akin to bulk $RE$NiO$_3$. In sharp  contrast, for  2ENO/1LNO SL the period of two kinds of inequivalent nickels [Ni(I), Ni(II)] is  $3\times c$ along [0 0 1]   and clearly cannot be  matched with the periodicity of the bulk-like checker board CO ($2 \times c$). Though both samples show  the first-order MITs and antiferromagnetism,  resonant X-ray scattering (RXS) at the Ni $K$-edge  revealed an unaffected amount of CO across the MIT for 2ENO/1LNO SL whereas the MIT of 1ENO/1LNO was accompanied by a significant modulation of CO. Combined magnetic characterizations using X-ray spectroscopy and scattering at Ni $L_{3,2}$-edges confirm  $S$ =1  behavior, signifying the critical importance of ligand holes for the magnetic ordering of these materials.  The decoupling of the structural and electronic degrees of freedom and confirmation of $S$=1 magnetic nature attest to the fact that structural and CO transitions are not the origin of MIT for 2ENO/1LNO, and  selects the SSM scenario as the microscopic mechanism  for the transition~\cite{khomskii_holeordering,millis_siteselective,sawatzky_hf,georges_dmft,haule,green}.

[2ENO/1LNO]x12 and [1ENO/1LNO]x18  SLs were grown on orthorhombic NdGaO$_3$ (1 1 0)$_\mathrm{or}$ [(0 0 1)$_\mathrm{pc}$] (or and pc denote orthorhombic and pseudocubic settings respectively) substrates  by  laser MBE~\cite{derek_prb,derek_prb2,lno_optics,own_apl,scirep,own_prl}. The growth was monitored by in-situ reflection high energy electron diffraction. X-ray diffraction (XRD) measurements  (shown in supplemental~\cite{sup}), recorded  using the six-circle diffractometer at the 6-ID-B beam-line of the Advanced Photon Source (APS) at Argonne National Laboratory confirm high structural quality of the superlattices and also establish that all samples are single domain~\cite{derek_prb2,Upton}. Transport properties were measured in 4-probe  Van Der Pauw geometry with a Quantum Design physical property measurement system (PPMS).   X-ray absorption spectroscopy (XAS) and X-ray magnetic circular dichroism (XMCD) on the Ni $L_{2,3}$ edges  were performed at the 4-ID-C beam line of the APS. Magnetic structure was investigated at the resonant soft X-ray beamline  4.0.2  of the Advanced Light Source (ALS).

Fig. 2 shows the electronic and magnetic properties of the  SLs. As immediately seen, both 2ENO/1LNO and 1ENO/1LNO SLs are metallic at room temperature and  undergo a first order MIT  at 245 K and 155 K respectively with several orders of magnitude resistivity changes, similar to   bulk $RE$NiO$_3$ (upper panels of Fig. 2(a) and (b)). The large dissimilarity in $T_\mathrm{MIT}$, along with the difference in hysteric widths suggests a significant difference between these SLs. As for the magnetic order, E$^{\prime}$--AFM ordering of $RE$NiO$_3$   is characterized by the  (1/2, 0, 1/2)$_\mathrm{or}$ [(1/4, 1/4, 1/4)$_\mathrm{pc}$] magnetic wave vector, which can be viewed as a stacking of either $\uparrow \uparrow \downarrow \downarrow$ or $\uparrow \rightarrow \downarrow \leftarrow$ or $\uparrow0\downarrow0$  of pseudocubic (111) planes~\cite{own_review}. This  type of magnetism is rather robust and  has been found in bulk-like thick films, ultra-thin films and superlattices of $RE$NiO$_3$~\cite{staub_nnomag,keimer_raman,derek_nno,derek_prb2,sno,Frano_prl}. To  assess the magnetic structure of  our superlattices we performed   resonant magnetic X-ray scattering recorded at the Ni $L_3$  edge resonance energy of 852 eV.  The inset in Fig. 2(b) shows the presence of a strong (1/4, 1/4, 1/4)$_\mathrm{pc}$ reflection appearing below the transition temperature. The plot of  integrated intensity of the peak  as a function of $T$ (lower panel of Fig. 2(a)-2(b)) yields   the antiferromagnetic transition temperature of $T_N$=220 $\pm$ 5K for 2ENO/1LNO  and 155$\pm$5K for 1ENO/1LNO. A  direct comparison to  the transport data shows that for the 2ENO/1LNO SL there is  a finite separation between $T_N$ and $T_\mathrm{MIT}$  that has been additionally confirmed  by the magneto-transport measurements ~\cite{sup,SNO_Hall}.  These results demonstrate that by altering the ENO layer thickness within just one period of the superlattice, the temperature driven electronic and magnetic transitions can be made either simultaneous or separate.

 \begin{figure}
 \vspace{-0pt}
\includegraphics[width=.49\textwidth]{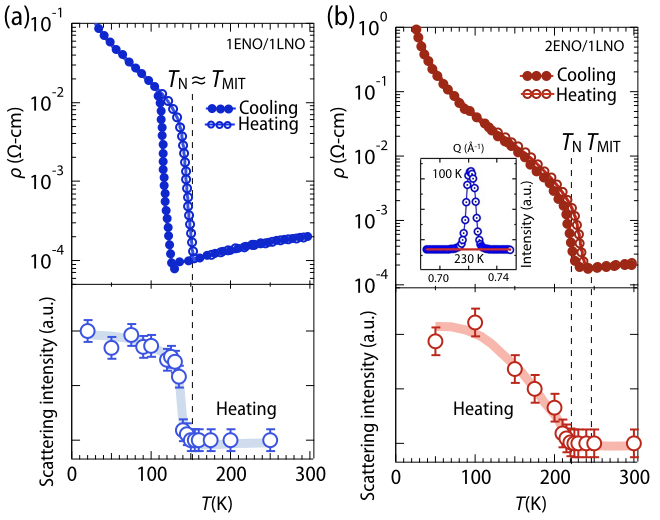}
\caption{\label{} Upper panel of (a) and (b) displays temperature dependence of the $dc$ resistivity of 1ENO/1LNO and 2ENO/1LNO respectively.  Lower panels show temperature dependence of the (1/2 0 1/2)$_\mathrm{or}$ Bragg peak  intensity corresponding to $E'$ type antiferromagnetic state. Magnified view of the resistivity around $T_\mathrm{MIT}$ has been shown in supplemental. Inset of (b) shows measured magnetic scattering for 100 K and 230 K for 2ENO/1LNO. }
\end{figure}

After establishing bulk-like magnetic order,   we investigate for a possible difference in CO, and the accompanying structural transition, as the progenitor of dissimilar transport behavior. To  probe for the presence of CO, resonant X-ray scattering (RXS) at the $K$-edge (1$s \rightarrow 4p$ transition)  around  ($h$ 0 $l$)$_\mathrm{or} $ and (0 $k$ $l$)$_\mathrm{or}$ reflections (odd $h$, $k$, $l$)  has been extensively used as a direct proof of a CO phase ~\cite{nno_co,scagnoliprb,lorenzoprb,keimer_scattering}. Specifically, we performed our RXS measurements on the Ni $K$-edge (8.34 keV) at the (0 1 1)$_\mathrm{or}$ reflection. The structure factor for this reflection can be written as~\cite{nno_co,scagnoliprb,lorenzoprb,keimer_scattering} $F_{011}(\textbf{Q},E) = A_{\textrm{O},{RE}}(\textbf{Q})+2\Delta f^{0}_\mathrm{Ni}(\textbf{Q})+2 \Delta f^{\prime}_{\textrm{Ni}}(E)+2 \Delta f^{\prime\prime}_{\textrm{Ni}}(E)$ with $\Delta f^{0}_\mathrm{Ni}(\textbf{Q})$=$f^{0}_{\textrm{Ni1}}(\textbf{Q})-f^{0}_{\textrm{Ni2}}(\textbf{Q})$, $\Delta f^{\prime}_{\textrm{Ni}}(E)=f^{\prime}_{\textrm{Ni1}}(E)-f^{\prime}_{\textrm{Ni2}}(E)$, $\Delta f^{\prime\prime}_{\textrm{Ni}}(E)=f^{\prime\prime}_{\textrm{Ni1}}(E)-f^{\prime\prime}_{\textrm{Ni2}}(E)$.  $A_{\textrm{O},{RE}}(\textbf{Q}),  f^{0}_{\textrm{Ni}}(\textbf{Q})$ are the energy independent Thompson scattering terms  for the $RE$ and O-sites, and the Ni-sites respectively. The $ f^{\prime}_{\textrm{Ni}}(E), f^{\prime\prime}_{\textrm{Ni}}(E)$ terms represent the real and imaginary energy-dispersive correction factors which contribute to the resonant behavior. The intensity of the Bragg reflection is then given by $I_{011} \propto \lvert F_{011}\rvert^2 = A_{\textrm{O},{RE}}^2+2A_{\textrm{O},{RE}}\cdot2(\Delta f^{0}_{\textrm{Ni}}+\Delta f^{\prime}_{\textrm{Ni}}+\Delta f^{\prime\prime}_{\textrm{Ni}}) +4(\Delta f^{0}_{\textrm{Ni}}+\Delta f^{\prime}_{\textrm{Ni}}+\Delta f^{\prime\prime}_{\textrm{Ni}})^2$.  As clearly seen,  because of the mixing of $A_{\textrm{O},{RE}}$  and $\Delta f$ parameters, any energy-dependent features at resonance can be due to either a change in the energy dispersive terms themselves, or due to a change in the $Q$-dependent Thompson scattering factors from the $RE$ and O-sites, $A_{\textrm{O},{RE}}(\textbf{Q})$. Thus, through this entanglement, a significant modulation of  $A_{\textrm{O},{RE}}$ may contribute to a large energy-dependent variation at resonance in addition to the Ni charge ordering. However,  all energy-dispersive terms  go to zero [$\Delta f^{\prime}_{\textrm{Ni}}(E)$ and $\Delta f^{\prime\prime}_{\textrm{Ni}}(E) \rightarrow 0$] a few eV away from the resonant energy,  and thus isolate changes due to the energy independent terms $A_{\textrm{O},{RE}}$ alone.

\begin{figure}
 \vspace{-0pt}
\includegraphics[width=.47\textwidth]{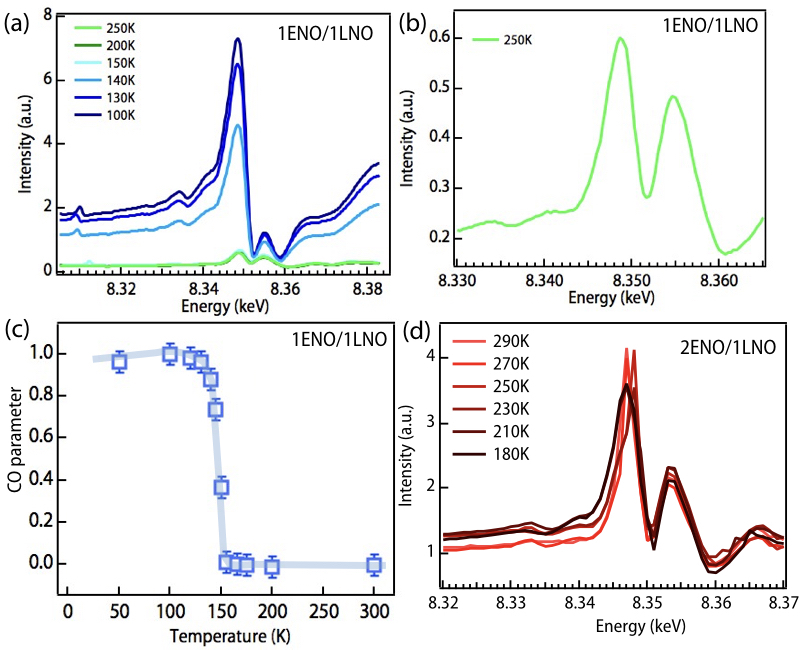}
\caption{\label{} (a) (0 1 1)$_\mathrm{or}$ resonance energy scan for the 1ENO/1LNO SL at various temperature. Resonance does also persist in the metallic state (b). (c) Temperature dependence of CO parameter ~\cite{coorderparameter} for 1ENO/1LNO.  (d) (0 1 1)$_\mathrm{or}$ resonance for the 2ENO/1LNO SL showing a strong resonance signal but no significant change across $T_\mathrm{MIT}\sim$ 245K is observed. }
 \end{figure}

The 1ENO/1LNO SL  exhibits a sharp appearance  of the (0 1 1)$_\mathrm{or}$ reflection below the MIT as exemplified by the data at 100K shown in Fig. 3(a). The result is similar  to what is seen in bulk-like  films of NdNiO$_3$ and has been directly connect to the monoclinic CO phase~\cite{nno_co,scagnoliprb,lorenzoprb,keimer_scattering}.   Upon heating across the MIT, the CO scattering signal from 1ENO/1LNO SL is strongly reduced, but  in contrast to the expectations for the bulk nickelates~\cite{nno_co}, does not completely disappear, and is likely due to the presence of octahedral distortions contributing to the off-diagonal elements of the energy dispersive scattering factors~ \cite{keimer_scattering}. The large change in the off-resonant  scattering intensity for 1ENO/1LNO  emphasizes a significant lattice modulation across the MIT.  The coupling of the energy-independent and dispersive factors can also affect the resonance intensity and shape, and thus  contributes to the strong change across the MIT observed at resonance.  The  change in CO is quantified in Fig. 3(c) where we plot  $T$-dependence  of the charge order parameter, defined as $\alpha =\sqrt {I (E_\mathrm{res})}$ -$\sqrt {I(E_\mathrm{offres})}$~\cite{nno_co,coorderparameter}, which is concurrent  with both the MIT and antiferromagnetic order present in 1ENO/1LNO SL.

The case of 2ENO/1LNO shown in  Fig. 3(d)  shows a stark contrast to 1ENO/1LNO. Specifically, no significant change in either the line shape or intensity occurs upon traversing the MIT at $\sim$ 245 K for 2ENO/1LNO SL.  Moreover, no change in the on- or off-resonant behavior unambiguously implies that there is no significant modulation of the structure factor across the MIT for 2ENO/1LNO, which in turn  \textit{excludes} the presence of bulk-like long-range CO and  lattice symmetry change within the detection limit of the scattering technique.  However, short range CO could still be present~\cite{srangeco}. The lack of bulk-like CO or structural symmetry change across the MIT means another driving mechanism is at play. Additionally, the large resonance intensity found above $T_\mathrm{MIT}$ for 2ENO/1LNO SL signifies the realization of \text{a new monoclinic metallic phase} without  application of any external field.

Next,  we focus on a possible  microscopic mechanism responsible for the MIT. In contrast to 1ENO/1LNO SL where the simultaneous transitions can be accounted by a spin-density wave scenario~\cite{balents_landau,balents_hf,Frano_prl}, the observation of a MIT and magnetic transition in this 2ENO/1LNO SL without CO advocates for a different model such as the  site-selective Mott  mechanism~\cite{millis_siteselective}.  The key  aspect of the model  is that, in contrast to what is expected  from a Ni${^3+}$ picture (spin $S$=1/2  state),   an $S$=1-like magnetic ground state is predicted. Moreover, as very recently demonstrated in the SSM~\cite{haule} model, the Ni $K$-edge resonance line shape can be successfully generated from inequivalent Ni-sublattices with BD and  does not require any real CO among the Ni sites. To  test  such scenario for 2ENO/1LNO SL, we investigated the magnetic state of the Ni spins by carrying out XMCD measurements which is a local probe of element specific magnetism. Although the antiferromagnetic arrangement of Ni sites does not have any net magnetization, an applied external magnetic field ($H$) can cant the individual spins resulting  in a finite XMCD signal. If the applied magnetic field  $H$ energy is weaker than the crystal field, it preserves the intrinsic spin value and thus provides an opportunity to probe the magnitude of spin per nickel site. Figure 4(a) shows the X-ray  absorption  and XMCD  spectra at 50 K under a applied  field of 5 Tesla. For  comparison, we show the XAS of a well-known S=1 compound Ni$^{2+}$O . Surprisingly, the most notable feature of the XMCD line shape of the $L_3$ edge is that the energy corresponding to the (negative) peak of XMCD is at the same position as the peak of Ni$^{2+}$ XAS. Moreover, the XMCD line shape of 2ENO/1LNO is strikingly similar to the XMCD of Ca$_2$NiOsO$_6$ (Ni$^{2+}$ with octahedral crystal field analogous to $RE$NiO$_3$) obtained from Ref.~\cite{nixmcd} (see~\cite{sup}).  All these similarities point towards the existence of the $S$=1 state in 2ENO/1LNO SL~\cite{rixs}.

    \begin{figure}
 \vspace{-0pt}
\includegraphics[width=.45\textwidth]{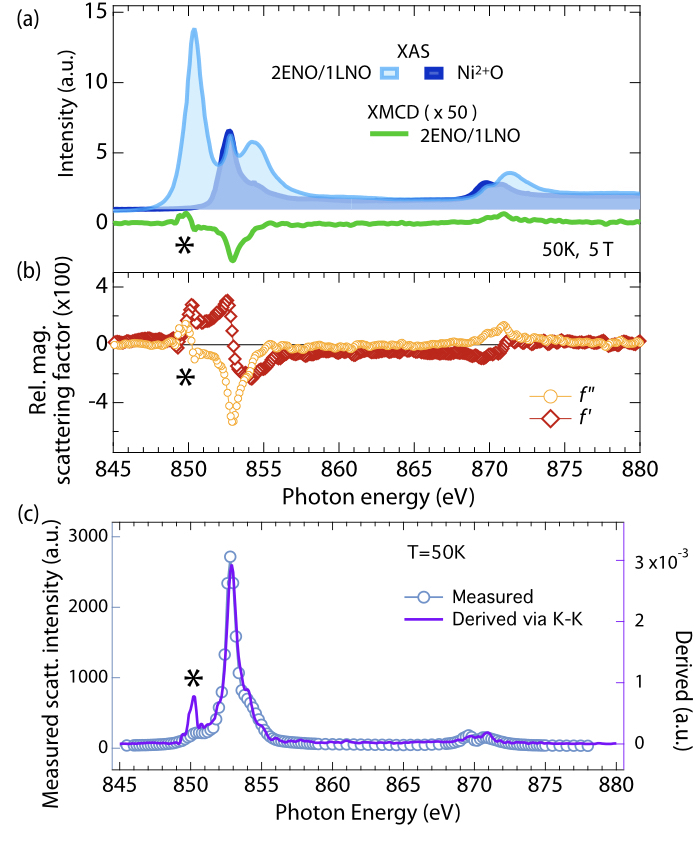}
\caption{\label{} (Color online) (a) XAS and XMCD of 2ENO/1LNO SL, measured with $H$ = 5 Tesla at 50 K and XAS of NiO. To rule out any artifacts, XMCD was measured with both +5 T and -5 T  and  their difference (divided by 2) has been plotted.  The strong artifacts in XMCD at the very strong La $M_4$ peak position does not exactly cancel with this field flipping, resulting in the small artifacts indicated by *. (b) Real and imaginary part of magnetic scattering factor. (c)  Energy dependence of magnetic scattering intensity was simulated using the data shown in (b) and is compared with the experimentally observed energy scan of  $E'$ magnetic structure.} 
\end{figure}

To investigate  whether $S$=1 like behavior of the film deduced from XMCD corresponds to the observed E$^{\prime}$-AFM,  we have derived magnetic scattering (Fig. 4(b)) from  XMCD data ~\cite{scaxmcd}  by using Kramers-Kronig transformation~\cite{scatterindxmcd1,scatterindxmcd2} and  compared it with the experimentally observed magnetic scattering data corresponding to the (1/4 1/4 1/4)$_\mathrm{pc}$ reflection.  As clearly seen in Fig. 4(c), the derived  magnetic scattering intensity has  exactly the same line shape and line positions as the energy scans recorded  on (1/4 1/4 1/4)$_\mathrm{pc}$ reflections  in our RXS experiment  which in turn  unambiguously confirms the $S$ = 1 magnetic state of Ni. To  emphasize, the MIT in 2ENO/1LNO without any bulk-like symmetry change,  the absence of CO, and  $S=1$-like behavior with $E^\prime$-AFM provide the first  experimental evidence for the theoretically proposed SSM ground state~\cite{millis_siteselective,sawatzky_hf,khomskii_holeordering,georges_dmft,haule,green}. Interestingly, the temperature dependence of out-of-plane lattice constant shows almost negligible lattice expansion ($\sim$0.18\%) across the MIT~\cite{sup}, which can be  also accounted by the bond disproportionation in SSM model as the partial volume compression of octahedra with $d^8\underline L^2$ configuration being compensated by the equal expansion of the neighboring octahedra with $d^8$ state~\cite{sawatzky_hf}.

In conclusion, by designing ultra-thin superlattices, we have been able to solve a long-standing puzzle of the MIT  in $RE$NiO$_3$ by demonstrating that neither charge ordering nor lattice symmetry transition is mandatory for the MIT, and  instead it can be accounted by a bond-disproportionation within the pure site-selective Mott  transition. The realization of a previously  unknown  monoclinic metallic phase without any external field demonstrates the utility of interface engineering as a tool for generating novel  materials phases of complex oxides. Moreover, the idea of selective suppression of cooperative ordering by mismatching the  structural periodicity  with the periodicity of the ordering phenomenon can be a promising route to unravel the mystery of competing phases including charge density wave and pseudo gap phase on superconductivity in high $T_c$ cuprates~\cite{kiemer_nature} and may eventually provide a new materials design principle to enhance  $T_c$.

 J.C., X.L. Y.C. are supported by the Gordon and Betty Moore Foundation EPiQS Initiative through Grant No. GBMF4534.  S.M.  is supported by IISc start up grant  and also  acknowledges Gordon and Betty Moore Foundation EPiQS Initiative for sponsoring a visit to Rutgers University. This research used resources of the Advanced Light Source, which is a Department of Energy Office of Science User Facility under Contract No. DE-AC0205CH11231. This research used resources of the Advanced Photon Source, a U.S. Department of Energy Office of Science User Facility operated by Argonne National Laboratory under Contract No. DE-AC02-06CH11357.


\begin{thebibliography}{99}
\bibitem{coorderparameter} CO($T$) for 1ENO/1LNO SL is defined as $\frac{\big(\sqrt {I (E_\mathrm{res})} -\sqrt {I(E_\mathrm{offres})}\big)_{T=T\mathrm K}-\big(\sqrt {I (E_\mathrm{res})} -\sqrt {I(E_\mathrm{offres})} \big)_{T=300\mathrm K}}{\big(\sqrt {I (E_\mathrm{res})} -\sqrt {I(E_\mathrm{offres})}\big)_{T=100\mathrm K}-\big(\sqrt {I (E_\mathrm{res})} -\sqrt {I(E_\mathrm{offres})} \big )_{T=300\mathrm K}}$.
\bibitem{scaxmcd}  In the long wavelength limit, total magnetic scattering cross section ($\sigma_m(E)$) is proportional to ($ \mid f_{m}'(E)\mid^2$+$\mid f_{m}'' (E)\mid^2$).
\end{thebibliography}
 \end{document}